\documentclass[11pt,oneside,letterpaper]{article}
\pdfoutput=1
\usepackage{amssymb}
\usepackage{amsmath}
\usepackage{graphicx}
\usepackage{setspace}
\usepackage{fancyhdr}
\usepackage{xcolor}
\usepackage{ifpdf}
\usepackage{graphicx}
\usepackage{rotating}
\usepackage{comment}
\usepackage{cite}

\usepackage{color}
\definecolor{darkgreen}{rgb}{0,0.5,0}
\definecolor{darkblue}{rgb}{0,0,0.6}
\definecolor{purple}{rgb}{0.4,.2,0.7}
\definecolor{awesome}{rgb}{1.0, 0.13, 0.32}

\usepackage[colorlinks=true,citecolor=darkgreen,linkcolor=purple,urlcolor=purple]{hyperref}

\def\d{\delta}

\newcommand{\bi}{\begin{itemize}}
\newcommand{\ei}{\end{itemize}}
\newcommand{\bea}{\begin{eqnarray}}
\newcommand{\eea}{\end{eqnarray}}
\newcommand{\be}{\begin{equation}}
\newcommand{\ee}{\end{equation}}

\definecolor{egyptianblue}{rgb}{0.06, 0.2, 0.65}

\numberwithin{equation}{section}
\allowdisplaybreaks  

\begin{document}
\begin{flushright}
{\tt MIT-CTP-4774}
\end{flushright}

\vspace*{2.5cm}
\begin{center}
{ \LARGE \textsc{Disordered Quivers and Cold Horizons}}
\\ 
\vspace*{1.7cm}
Dionysios Anninos$^\triangle$, Tarek Anous$^\#$ and Frederik Denef$^{\dag}$ \\
\vspace*{0.6cm}
\vspace*{0.6cm}
\it{\footnotesize $^\triangle$ School of Natural Sciences, Institute for Advanced Study, Princeton, NJ 08540, USA} \\
\it{\footnotesize $^\#$ Center for Theoretical Physics, Massachusetts Institute of Technology, Cambridge, MA 02139, USA} \\
\it{\footnotesize $^\dag$ Department of Physics, Columbia University, 538 West 120th Street, New York, New York 10027 }\\

\vspace*{0.6cm}

\vspace*{0.6cm}


\end{center}
\vspace*{1.5cm}
\begin{abstract}
\noindent

We analyze the low temperature structure of a supersymmetric quiver quantum mechanics with randomized superpotential coefficients, treating them as quenched disorder. These theories describe features of the low energy dynamics of wrapped branes, which in large number backreact into extremal black holes. We show that the low temperature theory, in the limit of a large number of bifundamentals, exhibits a time reparametrization symmetry as well as a specific heat linear in the temperature. Both these features resemble the behavior of black hole horizons in the zero temperature limit. We demonstrate similarities between the low temperature physics of the random quiver model and a theory of large $N$ free fermions with random masses.

\end{abstract}

\newpage
\setcounter{page}{1}
\pagenumbering{arabic}

\tableofcontents
\setcounter{tocdepth}{2}

\onehalfspacing

\section{Introduction}

In this work we consider how certain techniques developed in the study of systems with quenched disorder may be used in the context of string theory. The rough picture goes as follows: consider wrapping branes on complicated cycles within some compact manifold such that the branes are point-like in the non-compact space. These internal cycles can intersect amongst each other and amongst themselves, giving rise to a large number of light degrees of freedom localized at the brane intersections. Moreover, the different intersection modes can further interact with each other via some stringy processes. If the charges of the point-like branes in the non-compact directions, which are given by the cohomology of the cycles they wrap, become large enough we expect them to backreact into a charged extremal black hole. We expect several of the features of such black holes to be governed by the effective quantum mechanical theory obtained upon reducing the wrapped cycles to the lowest Kaluza-Klein modes. Generically this will be governed by some rather complicated Hamiltonian, coupling the large number of intersection modes between each other. The parameters of this Hamiltonian will result from difficult calculations involving the detailed structure of the compactification manifold. But perhaps, to some good approximation, they can simply be treated as a collection of random variables. If, in addition, these random variables evolve on time scales much larger than that of the intersection modes themselves, they may be viewed as quenched disorder. Our interest is to analyze the effect of this type of quenched disorder in a particular quantum mechanical model motivated from string theory. This approach is similar in spirit to Wigner's \cite{wigner}, who analyzed the spectral properties of heavy nuclei by approximating their Hamiltonian by a random matrix.

The model we consider is a supersymmetric quantum mechanics with four supercharges \cite{Douglas:1996sw,Denef:2002ru}. The matter content is organized in a quiver diagram, with bifundamental fields connecting different nodes and adjoint fields residing on the nodes. These models have been argued to describe, at weak coupling, the low energy physics of branes wrapping different cycles in a Calabi-Yau compactification. The adjoint matter describes the degrees of freedom of the brane's motion in the non-compact directions, and the bifundamentals capture the intersection modes in the internal cycles. Interestingly, certain types of quivers exhibit an exponentially large number of ground states \cite{Denef:2007vg} (in the large intersection number limit) reminiscent of the Bekenstein-Hawking entropy of the extremal black holes they are supposed to describe at strong coupling. The Hamiltonian of these models is highly constrained by supersymmetry, but it does allow for a superpotential which leads to interactions between the different intersection modes. The coefficients of this superpotential are in principle fixed by an often prohibitively difficult calculation depending on many of the details of the compactification manifold and its internal cycles. It is these superpotential coefficients that we take to be random in this paper, as a first step toward our broader goal.

The extremal black holes that the branes backreact into have several features of interest such as an AdS$_2 \times S^2$ near horizon with an $SL(2,\mathbb{R})$ symmetry and the possible fragmentation of this throat into a multitude of split horizons \cite{majumdar,papapetrou,Denef:2000nb,Anninos:2011vn,Anninos:2013mfa}. It was previously shown that the classically chaotic \cite{Anninos:2012gk} quiver models on the Coulomb branch, i.e. when the branes are separated in the non-compact space, are described in the  low energy limit by an $SL(2,\mathbb{R})$ invariant multi-particle mechanics \cite{Anninos:2013nra}. This is reminiscent of the $SL(2,\mathbb{R})$ invariant mechanics \cite{fe,Britto-Pacumio:2001zma} defined by the motion of the tips in the fragmented extremal throat.

We focus instead on the Higgs branch of the model, where the branes are sitting on top of each other and the intersection modes are light. In order to analyze this branch, we invoke the replica trick which involves considering $n$ replicas of the original system. Upon integrating out the random disorder, these replicas interact amongst each other. At large intersection number $N$, the system is described by a collection of $n\times n$ replica matrices $Q_{AB}$. The `paramagnetic' case $Q_{AB} = Q \, \delta_{AB}$ is shown to be perturbatively stable. Moreover, as speculated in \cite{Anninos:2013nra}, we find that the disorder averaged theory exhibits an emergent time reparametrization symmetry at low temperatures, containing an $SL(2,\mathbb{R})$ subgroup. This is reminiscent of the symmetry found in the near horizon region of extremal black holes \cite{Strominger:1998yg}. This is very similar to the situation encountered in several models of quenched quantum systems, such as the system recently considered in \cite{sachdev,parcollet,kitaev,Sachdev:2015efa,Polchinski:2016xgd,maldacena}. Additionally, we study the low temperature thermodynamics of the system. We show that the specific heat at low temperatures grows linearly. Interestingly, the specific heat of near extremal black holes has a similar linear growth in $T$. In appendix \ref{fermionM} we discuss how a simple model of free fermions with random masses also exhibits several of these features.

Finally, we establish that the Coulomb branch and Higgs branch $SL(2,\mathbb{R})$ invariant sectors are distinct---the scaling dimensions of the fields in each sector is different. We also initiate the study of more general replica symmetric and replica symmetry breaking saddles, but leave a complete analysis of this question to future work.


\section{Quiver quantum mechanics}\label{quiverintro}

The quantum mechanical theories of interest in this paper have four supercharges \cite{Denef:2002ru}. The matter content resides in a chiral multiplet $\Phi^i_\alpha = (\phi^i_\alpha,\psi^i_\alpha,F^i_\alpha)$ containing a complex scalar $\phi^i_\alpha$, a complex Weyl spinor\footnote{The Weyl spinor has an $SO(3)$ spinor index which we are suppressing. The $\psi_\alpha^i$ transform in the $\bold{2}$ of $SO(3)$ and  $\bar{\psi}_\alpha^i$ transform in the $\bar{\bold{2}}$.} $\psi^i_\alpha$ and an auxiliary complex scalar $F^i_\alpha$. The index $\alpha = 1,2,\ldots,N$ indicates a particular intersection mode connecting two branes, and the index $i=1,2,3,\ldots,m$ represents the particular pair of branes being connected. We consider a cyclic quiver with three nodes (and hence three branes), i.e. $m=3$, since it is the simplest case and the general $m$-node cyclic case turns out to be qualitatively similar. The Euclidean action contains a standard kinetic piece:
\begin{equation}\label{skin}
S_{\rm kin} =  \int d\tau \left( |\dot{\phi}^i_\alpha|^2  +  \bar{\psi}^i_\alpha  \dot{\psi}^i_\alpha  - |F^i_\alpha|^2 \right)~,
\end{equation}
as well as interactions governed by a holomorphic superpotential $W(\phi)$:
\begin{equation} 
S_{\rm int} =  \int d\tau  \left( \frac{\partial W(\phi)}{\partial \phi^i_\alpha } F^i_\alpha +  \frac{1}{2}  \frac{\partial^2 W(\phi)}{\partial \phi^i_\alpha \partial \phi^j_\beta} \psi^i_\alpha \epsilon \, \psi^j_\beta + h.c. \right)~.
\end{equation}
Repeated indices are summed throughout our discussion unless otherwise specified. The theory has an $SO(3)$ symmetry generated by $\bold{J} = \bar{\psi}^i_\alpha \boldsymbol{\sigma}\psi_\alpha^i / 2$. Notice that the above expression also contains an $SO(3)$ invariant term $\psi^i_\alpha \epsilon \, \psi^j_\beta$ containing the $2\times 2$ $\epsilon$-tensor which is contracted by the (suppressed) spinor indices of the fermions. The supersymmetry transformations act as follows:
\begin{eqnarray}
\delta \phi^i_\alpha &=& \sqrt{2} \, \xi \,  \epsilon \, \psi^i_\alpha~, \\
\delta \psi^i_\alpha &=&   \sqrt{2} \, \bar{\xi} \, \epsilon \, \dot{\phi}^i_\alpha  + \sqrt{2} \, \xi \, F^i_\alpha~, \\
\delta F^i_\alpha &=& \sqrt{2} \, \bar{\xi} \, \dot{\psi}^i_\alpha~.
\end{eqnarray}
We will work with a specific holomorphic superpotential:
{\begin{equation}
W(\phi) = \Omega_{\vec{\alpha}} \, \phi^1_{\alpha}  \phi^2_{\beta} \phi^3_{\gamma}~,
\end{equation}}
with $\vec{\alpha} \equiv (\alpha,\beta,\gamma)$ and {$\Omega_{\vec{\alpha}}$} a set of constants.{Since $W(\phi)$ contains no quadratic piece, the bosons $\phi^i_\alpha$ and fermions $\psi^i_\alpha$ are all massless.} There can be higher order terms, but we will only consider the lowest order one. The scalar potential is given by: 
\begin{equation}
V(\phi) = \sum_{\alpha, i}\left|\frac{\partial W(\phi)}{\partial \phi^i_\alpha} \right|^2~.
\end{equation}
 It is useful to provide the engineering dimensions of the various fields. With $[\tau] = +1$ we have {$[\Omega_{\vec{\alpha}}] = -3/2$}, $[\phi_\alpha] = +1/2$, $[\psi_\alpha] = 0$ and $[F_\alpha] = -1/2$. {Note that the $\Omega_{\vec{\alpha}}$ are the only parameters in the model, and they are dimensionful.} For the system at finite temperature $T = 1/\beta$ we identify $\tau \sim \tau+\beta$ with $[\beta] = +1$. 

As mentioned in the introduction, these theories can be viewed as low energy effective actions arising from the dynamics of open strings living at the intersection points of branes wrapped along internal cycles in the compactification manifold. The branes are point like in the non-compact dimensions, and we are discarding their position degrees of freedom (which comprise a vector multiplet) by making them parametrically massive. A more complete analysis would include these degrees of freedom. We briefly discuss their effect in section \ref{Coulomb}. For large values of $N$ the point-like branes can backreact into extremal black holes with an AdS$_2\times S^2$ throat in the near horizon geometry. The details of the coefficients {$\Omega_{\vec{\alpha}}$} are contained in the geometry of the compactification manifold and the specific cycles wrapped by the branes.


It is in general rather complicated to compute the exact values of {$\Omega_{\vec{\alpha}}$}. Therein lies the basic assumption of our paper. We take the {$\Omega_{\vec{\alpha}}$} to be random coefficients drawn independently from a Gaussian probability distribution with variance {$\langle |\Omega_{\vec{\alpha}}|^2 \rangle = \Omega^2$} and vanishing mean. Moreover we assume that the disorder is quenched, such that we must average over the disorder only upon computing a particular extensive, physical quantity such as the free energy. In what follows we analyze the implications of such quenched disorder.


{The supersymmetric ground state sector of these models has been the subject of extensive work \cite{Denef:2007vg,Manschot:2012rx,Bena:2012hf,Lee:2012sc}. By evaluating a Witten index, the particular model under consideration was shown to have an exact ground state degeneracy that grows as $2^{3N}$ in the large $N$ limit \cite{Denef:2007vg}. Our main interest in what follows regards the low temperature non-supersymmetric sector of the model which has been far less explored.}


\subsection{Partition function and the replica trick}

In order to compute the two-point function we consider the Euclidean partition function at finite temperature:\footnote{From now on we normalize all fields and parameters in units where $\beta = 1$. For example, we denote the dimensionless quantity $\Omega^2 \beta^3$ by $\Omega^2$ and reinstate factors of $\beta$ when necessary.}
{
\begin{equation}
Z_\Omega [T] = \int \mathcal{D} \Phi_\mathcal{I} \, e^{-S_E[\Phi_\mathcal{I};\, \Omega_{\vec{\alpha}}]}~,
\end{equation}}
where $\mathcal{I}$ is a generalized index specifying any of the given fields. The bosonic degrees of freedom are periodic around the thermal circle $\tau \sim \tau+1$ and the fermonic degrees of freedom are anti-periodic. For a given realization of disorder the above partition function is too complicated to analyze. However for large enough systems, disorder-averaged quantities get at the essential physics. Since the disorder is quenched, the main quantity of interest is the averaged free energy, given by the logarithm of {$Z_\Omega[T]$}. We express this as (with overlines denoting disorder averages):
{\begin{equation}
\overline{F_{\Omega}[T]}=-T\,\overline{\log Z_\Omega [T]} = -T\lim_{n \to 0} \partial_n  \overline{\left(Z_\Omega[T]\right)^n}~.
\end{equation}}
At this point take $n$ to be an integer, such that we can view {$(Z_\Omega[T])^n$} as $n$-replicas of the original system. The average over the quenched disorder is performed over this replicated system, and the basic assumption is that the final results can be analytically continued to real $n$. This assumption is known in the literature as the replica trick, and is a basic tool in analyzing simple models of spin glasses. As a warm up example in using the replica trick, we provide a simple solvable case of free fermions with a random mass matrix in appendix \ref{fermionM}.

Upon integrating out the disorder we find an effective action for the replicated degrees of freedom $\Phi^{i}_{\alpha A} = (\phi^i_{\alpha A},\psi^i_{\alpha A},F^i_{\alpha A})$ which now cary an additional replica index $A=1,\ldots,n$. For our choice of superpotential, this effective action reads:
\begin{equation}\label{seffdisorder}
S_{\rm eff} = S_{\rm kin} - \Omega^2   \int d\tau \, d\tau' \mathcal{F}_{\vec{\alpha}}[\Phi^i_{\alpha A}(\tau)] \bar{\mathcal{F}}_{\vec{\alpha}}[\bar{\Phi}_{\alpha A}^i(\tau')]~,
\end{equation}
where
\begin{equation}
\mathcal{F}_{\vec{\alpha}}[\Phi^i_{\alpha A}(\tau)] \equiv  \sum_{ \vec{i} \in \mathcal{S}_3} \sum_A \left( \phi^{i_1}_{\alpha A} \phi^{i_{2}}_{\beta A} F^{i_3}_{\gamma A} + \psi^{i_1}_{\alpha A} \epsilon \psi^{i_2}_{\beta A}  \phi^{i_3}_{\gamma A} \right)~.
\end{equation}
$\mathcal{S}_3$ is the permutation group with $3$-elements and $\vec{i} = (i_1,i_2,i_3)$. The kinetic action $S_{\rm kin}$ is the replicated version of the original (\ref{skin}):
\begin{equation}\label{kinR}
S_{\rm kin} = \sum_{A,i} \int d\tau \left( -\bar{\phi}^i_{\alpha A} \partial^2_\tau \phi^i_{\alpha A}  + \bar{\psi}^i_{\alpha A} \partial_\tau \psi^i_{\alpha A} - \bar{F}^i_{\alpha A} F^i_{\alpha A} \right)~.
\end{equation}

Notice that upon integrating out the disorder we have coupled the replica indices. Also note that since the original action is bounded from below, the path integral over the replica matrices must be well defined for all $\Omega^2>0$. Furthermore, (\ref{seffdisorder}) is invariant under the same supersymmetry transformations as the original action. This is to be expected, since the theory is supersymmetric for any given realization of the variables $\omega_{{\vec{\alpha}}}$.

\subsection{Replica matrices}

At this point we introduce replica matrices: $Q_{AB}^{\mathcal{I}\mathcal{J}}(\tau,\tau')$ and delta-functionals implementing the on-shell conditions:
\begin{equation}
Q_{AB}^{\mathcal{I}\mathcal{J}}(\tau,\tau') =  \sum_{\alpha} \bar{\Phi}_{\alpha A}^{\mathcal{I}} (\tau) {\Phi}_{\alpha B}^{\mathcal{J}}(\tau')~,
\end{equation}
where we use the generalized index $\mathcal{I} = \{  i , \text{type} \}$ to specify the node index $i$ and the field type, i.e. $\phi$, $F$ or $\psi^a$. These are reminiscent of the bi-local fields introduced in \cite{Das:2003vw}. We use the integral representation of the delta functionals by introducing Lagrange multipliers $\Lambda^{\mathcal{I}\mathcal{J}}_{AB}(\tau,\tau')$:
\begin{multline}
\delta\left[ Q^{\mathcal{I}\mathcal{J}}_{AB}(\tau,\tau') - \bar{\Phi}^\mathcal{I}_{\alpha A} (\tau) {\Phi}^{\mathcal{J}}_{\alpha B}(\tau') \right] = \\ \int \mathcal{D}\Lambda^{\mathcal{I}\mathcal{J}}_{AB}(\tau,\tau') \, \exp\left[ i \int d\tau d\tau' \Lambda^{\mathcal{I}\mathcal{J}}_{AB}(\tau,\tau') \left( Q^{\mathcal{I}\mathcal{J}}_{AB}(\tau,\tau') - \bar{\Phi}^\mathcal{I}_{\alpha A} (\tau) \Phi^\mathcal{J}_{\alpha B}(\tau')  \right) \right]~.
\end{multline}
Upon implementing the delta function conditions, the remaining $S_{\rm eff}$ is quadratic in the $\Phi^{\mathcal{I}}_{\alpha A}$, which we can consequently integrate out. This leads to a Berezinian determinant factor and an interacting action in the bilocal fields. Writing (\ref{kinR}) as $S_{\rm kin} = \int d\tau d\tau' \bar{\Phi}_{\alpha A}^\mathcal{I}(\tau) \mathcal{O}^{\mathcal{I}\mathcal{J}}(\tau,\tau') \Phi_{\alpha A}^{\mathcal{J}}(\tau')$, the determinant factor reads:
\begin{equation}\label{Zeff}
\text{Det} \left[ \mathcal{O}^{\mathcal{I}\mathcal{J}}(\tau,\tau') \otimes \mathbb{I}_{n\times n} +  i \Lambda^{\mathcal{I}\mathcal{J}}_{AB}(\tau,\tau') \right]^{-N}~.
\end{equation}
If we had chosen to turn on a source $J_{A\alpha}^{\mathcal{I}}(\tau)$ for the $\Phi^{\mathcal{I}}_{A\alpha}(\tau)$ fields, the partition function would also be a function of these $J_{A\alpha}^{\mathcal{I}}(\tau)$. In this case integrating out the $\Phi^{\mathcal{I}}_{A\alpha}(\tau)$ leads to an additional term in the effective action:
\begin{equation}\label{source}
S_{\rm source}[{J}^{\mathcal{I}}_{A}(\tau)] = -\int d\tau d\tau' \, \bar{J}_{A\alpha}^{\mathcal{I}}(\tau) \left[ \mathcal{O}^{{\mathcal{I}\mathcal{J}}} \otimes \mathbb{I}_{n\times n} + i \, \Lambda^{\mathcal{I}\mathcal{J}}_{AB}(\tau,\tau') \right]^{-1} {J}_{B\alpha}^{\mathcal{J}}(\tau')~.
\end{equation}
The above expression expresses the correlation functions of the original fields $\Phi$ in terms of the new variables $Q$ and $\Lambda$.

\subsection{Large $N$ limit}

Notice that the functional determinant in (\ref{Zeff}) is raised to the power $N$. We are interested in a particular large $N$ limit, which we will now specify. We rescale $Q \to N {Q}$ and keep $\Omega N \equiv \lambda$, ${Q}$ and $\Lambda$ fixed as $N\to \infty$. With this large $N$ limit all exponents in the effective action scale as $N$.

It follows from (\ref{Zeff}) and (\ref{source}) that the saddle point value of the $Q^{\mathcal{I}\mathcal{J}}_{AB}(\tau,\tau')$ computes the various disorder averaged, equilibrium, two-point functions:
\begin{equation}
Q^{\mathcal{I}\mathcal{J}}_{AB}(\tau,\tau') = \frac{1}{N} \, \langle \bar{\Phi}^{\mathcal{I}}_{\alpha A}(\tau) {\Phi}^{\mathcal{J}}_{\alpha B} (\tau') \rangle~.
\end{equation}
 It is useful to note that several of these $Q^{\mathcal{I}\mathcal{J}}_{AB}(\tau,\tau')$ will vanish. (For example, they might involve a single fermionic field.) Moreover, there will always be a large $N$ saddle point for which the replica matrices $Q^{\mathcal{I}\mathcal{J}}_{AB}(\tau,\tau')$  all have $\mathcal{I} = \mathcal{J}$. This saddle will be perturbatively stable against fluctuations in the $\mathcal{I}\neq\mathcal{J}$ directions. We study such saddles in what follows, since they already contain a lot of interesting phenomena. 

In the $\mathcal{I} = \mathcal{J}$ subspace, the effective action simplifies considerably. Thus we can simplify our notation slightly. We denote:
\begin{eqnarray}
Q_{AB}^{i}(\tau,\tau') &\equiv& \frac{1}{N} \langle \bar{\phi}_{\alpha A}^i(\tau) \phi^i_{\alpha B}(\tau') \rangle~,\\
P_{AB}^{i}(\tau,\tau') &\equiv& \frac{1}{N}  \langle \bar{F}_{\alpha A}^i(\tau) F^i_{\alpha B}(\tau') \rangle~, \\
S_{AB}^{i,a}(\tau,\tau') &\equiv& \frac{1}{N}  \langle \bar{\psi}_{\alpha A}^{i,\dot{a}}(\tau) \psi^{i,a}_{\alpha B}(\tau')  \rangle~.
\end{eqnarray}
In the above expressions, only the Greek indices are summed over, and the index $a=1,2$ for the fermionic variables is the $SO(3)$ spinor index. Our effective action thus becomes:{
\begin{multline}\label{effectiveactionwP}
\frac{S_{\rm eff}}{N} =  \sum_{i=1}^3  \left[  \int d\tau d\tau'  \delta(\tau-\tau') \left( - \partial_\tau^2 Q_{AA}^i +\sum_{a=1}^2\partial_\tau S_{AA}^{i,a} - P_{AA}^i\right)   \right. \\ \left. -\text{tr} \log Q_{AB}^i(\tau,\tau')+ \sum_{a=1}^2 \text{tr} \log S^{i,a}_{AB}(\tau,\tau') -\text{tr}\log \left(-P_{AB}^i(\tau,\tau')\right) \right]  \\ -{\lambda^2} \sum_{\vec{i} \in \mathcal{S}_3} \int d\tau d\tau'Q_{AB}^{i_1}(\tau,\tau') \left(Q_{AB}^{i_2}(\tau,\tau')P_{AB}^{i_3}(\tau,\tau')+ S^{i_{2},1}_{AB}(\tau,\tau') S^{i_3,2}_{AB}(\tau,\tau')\right)~,
\end{multline}
where we have integrated out the Lagrange multipliers implementing the delta function constraints in deriving this action. The above action implies the following equations of motion for $P^i_{AB}(\tau,\tau')$:
\begin{multline}\label{Peq}
\delta_{AB}\delta(\tau-\tau')=P^i_{AB}(\tau-\tau')\\+\lambda^2\sum_C\int\d\tau''\,P^i_{CB}(\tau-\tau'-\tau'')\sum_{i\notin j\in\mathcal{S}^2}Q_{AC}^{j_1}(\tau'')Q_{AC}^{j_2}(\tau'')~.
\end{multline}
Since the kinetic term for $P^i_{AB}(\tau,\tau')$ has no time derivatives, we can integrate it out exactly, leading to the following action solely in terms of $Q^i_{AB}(\tau,\tau')$ and $S^{i,a}_{AB}(\tau,\tau')$:
\begin{multline}\label{effectiveaction}
\frac{S_{\rm eff}}{N} =  \sum_{i=1}^3  \left[  \int d\tau d\tau'  \delta(\tau-\tau') \left( - \partial_\tau^2 Q_{AA}^i + \sum_{a=1}^2\partial_\tau S_{AA}^{i,a} \right)  -\text{tr} \log Q_{AB}^i(\tau,\tau') \right. \\ \left. + \sum_{a=1}^2 \text{tr} \log S^{i,a}_{AB}(\tau,\tau') +  \sum_{k=1}^3\text{tr}\log \left( \mathbb{I}_{n\times n} \delta(\tau-\tau') + \lambda^2 \prod_{k\neq i}Q^{i}_{AB}(\tau,\tau') \right) \right]  \\ -{\lambda^2} \sum_{\vec{i} \in \mathcal{S}_3} \int d\tau d\tau' Q_{AB}^{i_1}(\tau,\tau') S^{i_{2},1}_{AB}(\tau,\tau') S^{i_3,2}_{AB}(\tau,\tau')~.
\end{multline}
 }

Since our theory is time translation invariant $Q^i_{AB}(\tau,\tau') = Q^i_{AB} (\tau - \tau')$ and $S^{i,a}_{AB}(\tau,\tau')  = S^{i,a}_{AB}(\tau-\tau')$.
Moreover, time reversal invariance implies for the diagonal components of the replica matrices: $Q_{AA}^i(\tau-\tau') = Q_{AA}^i(\tau'-\tau)$ which is moreover a real function. For the fermions, $S_{AA}^{i,a}(\tau-\tau') =  - S_{AA}^{i,a}(\tau'-\tau)$ is a real odd function of $u \equiv (\tau-\tau')$. In what follows, it will also be convenient to consider expressions in frequency space as well:
\begin{equation}
Q^i_{AB}(u) = \sum_{k\in\mathbb{Z}} e^{2\pi i k u}Q^i_{AB}(k)  ~, \quad S^{i,a}_{AB}(u) = \sum_{k\in\mathbb{Z}} e^{2\pi i (k+1/2) u} S^{i,a}_{AB}(k) ~.
\end{equation}
Note that $Q^i_{AB}(k) = \langle \bar{\phi}^i_{\alpha A}(k) \phi^i_{\alpha B}(-k) \rangle$ and $S^{i,a}_{AB}(k) = \langle \bar{\psi}^{i,a}_{\alpha A}(k) \psi^{i,a}_{\alpha B}(-k-1) \rangle$, where the $\phi^i_{\alpha A}(k)$ are the Fourier coefficients of $\phi^i_{\alpha A}(\tau)$ and so on. This in turn implies $Q_{AA}^i(k)$ is a real even function, and ${S}_{AA}^{i,a}(k) =  \overline{S_{AA}^{i,a}(-k-1)}$.

\section{Replica symmetry}\label{rep}

The action (\ref{effectiveaction}) is symmetric among the replica indices $A,B$. However, as is the case in spin glass models, replica symmetry can be spontaneously broken. In this section we discuss whether or not replica symmetry is broken in the model described above. To understand this question, one must understand the dominant contribution to the partition function at large $N$. In several simple models of quantum systems with quenched disorder \cite{braymoore,cugliandolo}, when replica symmetry is broken it indicates a transition to a glassy phase. A way to determine the presence of replica symmetry breaking is to study fluctuations about a replica symmetric saddle of the free energy and to  see whether it is locally stable.


 \subsection{Small fluctuations about the paramagnetic ansatz}

It follows from an argument in \cite{braymoore} that when $A \neq B$, the replica matrices must be time independent. The argument uses the fact that different replicas are decoupled before the disorder is integrated out. Hence the two-point function of two fields carrying different replica indices must factorize into a product of their respective one-point functions. However, the equilibrium value of the one-point function for a fixed realization of disorder is $\tau$-independent. From this it follows that $Q^i_{AB}$ with $A\neq B$ is itself $\tau$-independent, i.e. $Q^i_{AB}(\tau,\tau') = \left( \delta_{AB} Q^i(\tau,\tau') + q^i_{AB} \right)$ with $q^i_{AB} = \bar{q}^i_{BA}$ independent of $\tau$ and vanishing for $A=B$. For $S^{i,a}_{AB}(\tau,\tau')$, we have the even stronger condition that $S^{i,a}_{AB} = 0$ for $A \neq B$ since the equilibrium one-point function of a fermion $\langle \psi_{\alpha A}^{i,a} \rangle$ vanishes identically. The simplest and most symmetric ansatz sets the $q^i_{AB} = 0$, and we refer to this as the `paramagnetic' ansatz.

We would like to understand if this ansatz is a stable solution of the saddle point equations stemming from the large-$N$ action. The linear fluctuation of $S_{\rm eff}$ must vanish at the saddle point and thus the question is whether all quadratic contributions locally increase the value of the action. Expressing the fluctuations as $q^i_{AB}$ with $Q^i_{AB} = \delta_{AB} Q^i (\tau-\tau') + q^i_{AB}$ and expanding the effective action to second order in the fluctuations we find:
\begin{equation}
\frac{S^{(2)}_{\rm eff}}{N} =\frac{1}{2}\sum_{i}\int d\tau\,d\tau'\,\left((Q^i)^{-1}(\tau-\tau')\right)^2 \sum_{A,B}q^i_{AB}\bar{q}^i_{AB}~. 
\end{equation}
In the above expression $(Q^i)^{-1}$ is defined such that
\begin{equation}
\int d\tau_2 (Q^i)^{-1}(\tau-\tau_2)Q^i(\tau_2-\tau')=\delta(\tau-\tau')~.
\end{equation}
Notice that the quadratic fluctuation is positive definite. Thus, at least locally, the paramagnetic ansatz remains stable for all values of the temperature. This does not bar the possibility of a lower free energy configuration, only that such a configuration will be reached by non-perturbative fluctuations away from the paramagnetic one.

\subsection{Paramagnetic equations of motion}

Having argued that the paramagnetic ansatz is locally stable, we can examine the saddle point equations governing the diagonal elements of the replica matrix $Q_{AB}$ at large $N$. 
Taking the replica matrices to be diagonal, we are left with the following effective action:
\begin{multline}\label{parS1}
\frac{S_{\rm eff}}{Nn} = \sum_{i=1}^3\left[\int d\tau d\tau' \delta(\tau-\tau')\left(- \partial_\tau^2 Q^i + \sum_{a=1}^2\partial_\tau S^{i,a} \right)  -\text{tr} \log Q^i(\tau,\tau') \right.\\ \left.+ \sum_{a=1}^2\text{tr} \log S^{i,a}(\tau,\tau') +  \text{tr}\log \left(  \delta(\tau-\tau') + \lambda^2 \prod_{l\neq i}Q^{l}(\tau,\tau') \right)\right]  \\ -{\lambda^2} \sum_{\vec{i} \in \mathcal{S}_3} \int d\tau d\tau' Q^{i_1}(\tau,\tau') S^{i_{2},1}(\tau,\tau') S^{i_3,2}(\tau,\tau')~.
\end{multline}
From the above action, we find the following saddle equations:
\begin{align}
\delta(\tau-\tau') =&~ -\partial_\tau^2 Q^{i}(\tau-\tau')+\lambda^2\sum_{j\neq i}\sum_{k\in\mathbb{Z}}\int d\tau''\frac{Q^{i}(\tau-\tau'+\tau'')Q^{j}(\tau'')}{\int d\upsilon \,e^{2\pi i k(\upsilon-\tau'')}\left[\delta(\upsilon)+\lambda^2 Q^{i}(\upsilon)Q^{j}(\upsilon)\right]}\nonumber\\&-{\lambda^2}\int d\tau''Q^{i}(\tau-\tau'+\tau'')\sum_{i\notin \vec{j}\in \mathcal{S}_2} S^{j_1,1}(\tau'')S^{j_2,2}(\tau'')~, \label{saddle1}\\
\delta(\tau-\tau') =&~ -\partial_\tau S^{i,a}(\tau-\tau')\nonumber\\&+{\lambda^2}\int d\tau''S^{i,a}(\tau-\tau'+\tau'')\sum_{i\notin \vec{j}\in \mathcal{S}_2} Q^{j_1}(\tau'')S^{j_2,b}(\tau'') ~~~~~\text{for}~a\neq b~.  \label{saddle2}
\end{align}
These are the Schwinger-Dyson equations for the two-point functions of the scalar and fermion fields at large $N$.

We can also express the effective action and saddle point equations in frequency space:
\begin{multline}\label{parS}
\frac{S_{\rm eff}}{Nn} = \sum_{i=1}^3\sum_{k\in \mathbb{Z}}\left\lbrace(2\pi k)^2Q^i(k) + 2\pi i(k+1/2)\sum_{a=1}^2 S^{i,a}(k)  - \log Q^i(k)\right. \\ \left.+ \sum_{a=1}^2  \log S^{i,a}(k)+  \log \left[ 1 + \lambda^2 \prod_{l\neq i}\left(\sum_{k_l\in\mathbb{Z}}Q^{l}(k_l)\right)\,\delta\left(k+\sum_{j\neq i} k_j\right) \right]\right\rbrace   \\ -{\lambda^2} \sum_{\vec{i} \in \mathcal{S}_3} \sum_{k_1,k_2\in\mathbb{Z}} Q^{i_1}(k_{1})S^{i_{2},1}(k_{2}) S^{i_3,2}(-k_1-k_2-1)~,
\end{multline}
with the following saddle point equations:
\begin{align}
\frac{1}{Q^i(k)}=&~(2\pi k)^2+\lambda^2\sum_{j\neq i}\sum_{l\in\mathbb{Z}}\frac{Q^{j}(-k-l)}{1+\lambda^2 \sum_{m\in\mathbb{Z}}Q^i(m)Q^j(-m-l)}\nonumber \\
&-\lambda^2 \sum_{i\notin\vec{j} \in \mathcal{S}_{2}} \sum_{m\in\mathbb{Z}}S^{j_{1},1}(m) S^{j_{2},2}\left(-k-m-1\right)~,\label{bose2pt}\\
\frac{1}{S^{i,a}(k)}=&-2\pi i(k+1/2)
+{\lambda^2} \sum_{i\notin\vec{j} \in \mathcal{S}_{2}} \sum_{m\in\mathbb{Z}}Q^{j_1}\left(-k-m-1\right) S^{j_{2},b}(m) ~~\text{for}~a\neq b~.\label{fermi2pt}
\end{align}
When $\lambda=0$, it is easy to see that $S^{i,a}$ and $Q^i$ are precisely the frequency space two-point functions of a free fermion and a free boson respectively, and as expected.

We can study perturbative corrections to the free result in a small $\lambda$-expansion. These are given in appendix \ref{pert2} to first order. We find to leading order in $\lambda$ that the permutation symmetry acting on the node index remains unbroken. Also, $S^{i,a}(k)  = (2\pi i k)Q^i(k)$ (at large $k$) holds to sub-leading order in small $\lambda$ indicating that supersymmetry is preserved to this order. At large $\lambda$ the solutions may change significantly. We analyze this limit in the next section. Before doing so, we make some brief remarks about replica symmetry breaking.


\subsection{Replica symmetry breaking?}

Though the paramagnetic ansatz is perturbatively stable, we can still ask whether replica symmetry breaking takes place in this model. Though we do not analyze this question extensively in this work, we will make some brief comments in this section. First we need to study the simplest replica symmetric, yet non-paramagnetic, ansatz:
\begin{equation}\label{rsm}
Q_{AB}(k) =  Q(k) \, \delta_{AB}+  \upsilon \,  \epsilon_{AB} \,  \delta_{k,0}~, 
\end{equation}
where $\epsilon_{AB} \equiv  (1-\delta_{AB})$ is the matrix with vanishing diagonal terms and ones otherwise. We have dropped node and spinor indices, by assuming that in addition all the $Q^i_{AB}$ are equal amongst each other. The $\upsilon$ is a real constant independent of the frequency known as the Edwards-Anderson parameter in the spin glass literature \cite{ea}. 
 A non-vanishing Edwards-Anderson parameter often indicates the presence of glassy behavior.

With this ansatz, the effective action additionally becomes a function of $\upsilon$. The part that depends on the new variable $\upsilon$ is, for general replica size $n$:
\begin{multline}
\frac{S_{\rm eff}[\upsilon]}{3 N} = -(n-1) \log \left[ Q(0)-\upsilon \right]-\log\left[ (n-1) \upsilon +Q(0) \right] + \\ (n-1) \log \left[\lambda ^2 \left(Q\cdot Q-\upsilon^2\right)+1\right] +\log \left[\lambda ^2 \left((n-1) \upsilon^2+Q\cdot Q \right)+1\right]~, 
\end{multline}
where we define $Q \cdot Q \equiv \sum_m Q(m) Q(-m)$. The equation of motion for $\upsilon$ in the $n\to 0$ limit is given by:
\begin{equation}\label{eomfluct}
\frac{\upsilon}{[Q(0)-\upsilon]^2} = \frac{2 \lambda ^4 \upsilon^3}{\left[1+\lambda ^2 (Q \cdot Q-\upsilon^2)\right]^2}~.
\end{equation}
Clearly the paramagnetic value $\upsilon = 0$ is always a solution. To obtain the other solutions we must take into account the effect of non-vanishing $\upsilon$ on the equations governing $Q(k)$ and $S(k)$. This is a hard task, but inspection of (\ref{eomfluct}) already reveals that real solutions for $\upsilon$ are present given some $Q(k)$.
Assuming that there exists a solution with non-vanishing $\upsilon$, we can study fluctuations about this more general replica symmetric background. To do so, we consider quadratic perturbations $\delta Q_{AB}$ away from the replica symmetric background with a replica symmetry breaking perturbation. In perturbing away from the replica symmetric configuration we require $\delta Q_{AA} = 0$ and we also impose that $\text{tr} \, ( \epsilon \cdot \delta Q ) = \text{tr} \, (\epsilon \cdot \delta Q \cdot \epsilon \cdot \delta Q) = 0$. 
To obtain the quadratic action governing the perturbations, we must use the inverse of the replica symmetric matrices (\ref{rsm}) \cite{cugliandolo}:
\begin{equation}
Q^{-1}_{AB}(k) = \frac{1}{[Q(k)-\upsilon \, \delta_{k,0}]} \left(\delta_{AB} - \frac{\upsilon \, \delta_{k,0}}{[Q(k)+(n-1)\upsilon \, \delta_{k,0}]}\right)~.
\end{equation}
Also useful is the inverse:
\begin{multline}
\left[\mathbb{I}_{AB} + \lambda^2 \left( Q \cdot Q  \, \mathbb{I}_{AB} + \upsilon^2 \, \epsilon_{AB} \right)\right]^{-1} \\ = \frac{\left[\lambda ^2 \left(Q\cdot Q+(n-2) \upsilon^2\right)+1   \right]\delta_{AB}  - \upsilon^2 \lambda ^2   \epsilon_{AB}}{\left\lbrace\lambda^2 (\upsilon^2- Q\cdot Q)-1\right\rbrace\left\lbrace\lambda ^2 [(1-n)\upsilon^2- Q\cdot Q]-1\right\rbrace}~.
\end{multline}
The perturbations are dictated by the following action:
\begin{equation}
\frac{S^{(2)}}{N} =  \frac{3}{2}  A_Q \delta Q_{AB} \delta {Q}_{AB}~,  
\end{equation}
with (in the limit $n\rightarrow 0$):
\begin{equation}
A_Q = \frac{1}{\left[Q(0)-\upsilon\right]^2} - {(2\upsilon \lambda)^2}\left[{\frac{\lambda^2 \left(Q\cdot Q-2 \upsilon^2\right)+1}{\left(\lambda ^2 (\upsilon^2- Q\cdot Q)-1\right)^2} }\right]^2- \left[\frac{2\upsilon^2 \lambda ^4}{\left(\lambda ^2\left(\upsilon^2 - Q \cdot Q \right)-1\right)^2} \right]~. \\
\end{equation}
Using the equation of motion (\ref{eomfluct}), we note that the expression for $A_Q$ simplifies. The eigenvalue of the mass matrix of $\delta Q_{AB}$ is given by:
\begin{equation}
\lambda_Q =  -\frac{3}{2} {(2\upsilon \lambda)^2}\left[{\frac{\lambda^2 \left(Q\cdot Q-2 \upsilon^2\right)+1}{\left(\lambda ^2 (\upsilon^2- Q\cdot Q)-1\right)^2} }\right]^2 \le 0~.
\end{equation}
That $\lambda_Q$ is negative indicates that there is no stable replica symmetric saddle $\upsilon \neq 0$. It is always favorable to push in the direction of replica symmetry breaking!

What about these replica symmetry breaking saddles \cite{ea}? For instance, we could consider a single step replica symmetry breaking ansatz. This is given by splitting the $n$ replicas into $n/m$ clusters of size $m$. Within each $m\times m$ cluster the matrix takes the value $q_0$. For the pieces of the replica matrix not inside a given cluster, the replica matrix takes the value $\upsilon < q_1$. As before the diagonal components of the replica matrix take the value $Q(k)$. In other words:
\begin{equation}
Q_{AB}(k) = \left[ Q(k) - q_1 \delta_{k,0} \right]  \delta_{AB}  + \left[ q_1  - \upsilon \right] \, \epsilon^{(m)}_{AB}  \, \delta_{k,0}  + \upsilon \, \delta_{k,0} ~,
\end{equation}
where the matrix $\epsilon_{AB}^{(m)}$ is equal to one whenever $A$ and $B$ are within a diagonal $m\times m$ block. We leave it to future work to see whether such an ansatz, as well as the more general $k$-step replica symmetry broken matrices, lead to further stable large $N$ saddles.

\section{Scaling regime}

In this section we consider the theory in the large $\lambda$ limit. This can also be viewed as a low temperature limit, since upon reinstating the temperature $\lambda^2 \to \lambda^2 \beta^3$. Consider the saddle point equations (\ref{saddle1}) and (\ref{saddle2}). If it is the case that the terms proportional to $\lambda^2$ on the right hand side are dominant, then the theory exhibits a large symmetry:
\begin{eqnarray}\label{timerep}
\tau &\to& f(\tau)~, \\
 {Q}^{i}(\tau,\tau') &\to&
  \left( \frac{d f(\tau)}{d\tau} \, \frac{d f(\tau')}{d\tau'} \right)^{\mu_{Q^i}/2} \, {Q}^i\left(f(\tau),f(\tau') \right)~, \\
 {S}^{i,a}(\tau,\tau') &\to&
  \left( \frac{d f(\tau)}{d\tau} \, \frac{d f(\tau')}{d\tau'} \right)^{\mu_{S^{i,a}}/2} \, {S}^{i,a}\left(f(\tau),f(\tau') \right)~.
\end{eqnarray}
Moreover, it follows from (\ref{saddle1}) and (\ref{saddle2}) that the scaling coefficients must obey the following relation:
\begin{equation}\label{fullscaling}
\mu_{S^{i_1,1}} + \mu_{S^{i_2,2}} + \mu_{Q^{i_3}}   + 2 = 0~,  \quad  \vec{i} \in \mathcal{S}_3~.
\end{equation}
If the subgroup of the permutation symmetry permuting the $\mu_{Q_i}$ and $\mu_{S_{i}}$ remains unbroken, then $\mu_{Q_i} = \mu_Q$ and $\mu_{S^{i_1,a}} = \mu_S$ such that:
\begin{equation}\label{scaling}
2\mu_{S}+\mu_Q + 2 = 0~.
\end{equation}

We shall see in what follows that the above is indeed the saddle point solution at low temperatures, for given values of $\mu_Q$ and $\mu_S$. The above symmetry is a time reparametrization invariance, or in other words it is the set of diffeomorphisms (known as the  Witt algebra) that map the circle to itself. This vast symmetry group has a maximal finite dimensional sub-algebra generating the group $SL(2,\mathbb{R})$ of real $2\times 2$ matrices with unit determinant. Thus, the theory has an emergent conformal invariance in the particular scaling limit we have considered. We emphasize that for this symmetry to be precise, one requires a large $N$ and strong coupling limit.


\subsection{{Zero Temperature Solutions}}

We wish to find solutions to (\ref{bose2pt}) and (\ref{fermi2pt}) at strong coupling/low temperature. In order to make the low temperature limit more transparent, we will temporarily reintroduce factors of $\beta$ and take the $\beta\rightarrow\infty$ limit. To reintroduce units of $\beta$ we recall the discussion on units from Section \ref{quiverintro}{. Furthermore, with $\beta$ reintroduced, the thermal Fourier transform is:
\begin{equation}
f(k)=\frac{1}{\sqrt{\beta}}\int_0^\beta d\tau\,e^{i\omega_k\tau}f(\tau)~\,\quad\quad\quad f(\tau)=\frac{1}{\sqrt{\beta}}\sum_{k}e^{-i\omega_k\tau}f(k)~,
\end{equation}
  meaning that the units $ Q(k)$, and $S(k)$, differ from the units of $Q(\tau)$ and $S(\tau)$ by $[Q(k)]=[Q(\tau)]+1$ and similarly for $S(k)$ and $S(\tau)$.  In the $\beta\rightarrow\infty$ limit, the thermal frequencies $\omega_k=2\pi k/\beta$ become continuous and the sums over momenta can be replaced by integrals. Furthermore, since we have decompactified the thermal circle, there is no shift by one-half in the fermionic frequencies. We study the case where all $Q^i$'s and $S^{i,a}$'s are taken to be equal and drop the node and spinor $SO(3)$ indices entirely. The $\beta\rightarrow\infty$ limit yields:\footnote{
  {We remind the reader that the non-polynomial nature of (\ref{eomcont}) arises due to the fact that we have integrated out $P^i(\tau-\tau')$.
  }}
\begin{align}
\frac{1}{Q(\omega)}=&~\omega^2+2\lambda^2\int \frac{d\omega'}{2\pi}\frac{Q(-\omega-\omega')}{1+\lambda^2\int \frac{d\omega''}{2\pi}~Q(\omega'')Q(-\omega''-\omega')}\nonumber\\
&-2{\lambda^2} \int \frac{d\omega'}{2\pi} ~S(\omega') S\left(-\omega-\omega'\right)~,\label{eomcont}\\
\frac{1}{S(\omega)}=&-i\,\omega
+2{\lambda^2}\int\frac{d\omega'}{2\pi}  Q\left(-\omega-\omega'\right) S(\omega') ~.\label{eomcont2}
\end{align}
 Furthermore, at low energies ($\omega^3\ll \lambda^2$), we assume the following inequalities are satisfied:
\begin{equation}\label{inequality}
\omega^2 Q(\omega) \ll 1~, \quad  |\omega  S(\omega) | \ll 1~, \quad \int \frac{d\omega'}{2\pi}~Q(\omega')Q(-\omega-\omega') \gg \frac{1}{\lambda^2}~.
\end{equation}
 We will check that the solutions obtained under this assumption are indeed self-consistent.
Under our assumption, equations (\ref{eomcont}) and (\ref{eomcont2}) simplify:
\begin{align}
1=& ~2\int {d\omega'}\frac{Q(\omega)Q(-\omega-\omega')}{\int {d\omega''}~Q(\omega'')Q(-\omega''-\omega')}-2{\lambda^2} \int \frac{d\omega'}{2\pi} ~Q(\omega)S(\omega') S\left(-\omega-\omega'\right)~,\label{lowtemp1}\\
1=&~2{\lambda^2}\int\frac{d\omega'}{2\pi}  Q\left(-\omega-\omega'\right) S(\omega')S(\omega) ~.\label{lowtemp2}
\end{align}
Notice that the equations are self consistent. This can be seen by integrating both equations over $\omega$ and substituting (\ref{lowtemp2}) into (\ref{lowtemp1}).
}
\subsubsection{Non-supersymmetric solution}

\begin{figure}
\centering
\includegraphics[width=0.5\textwidth]{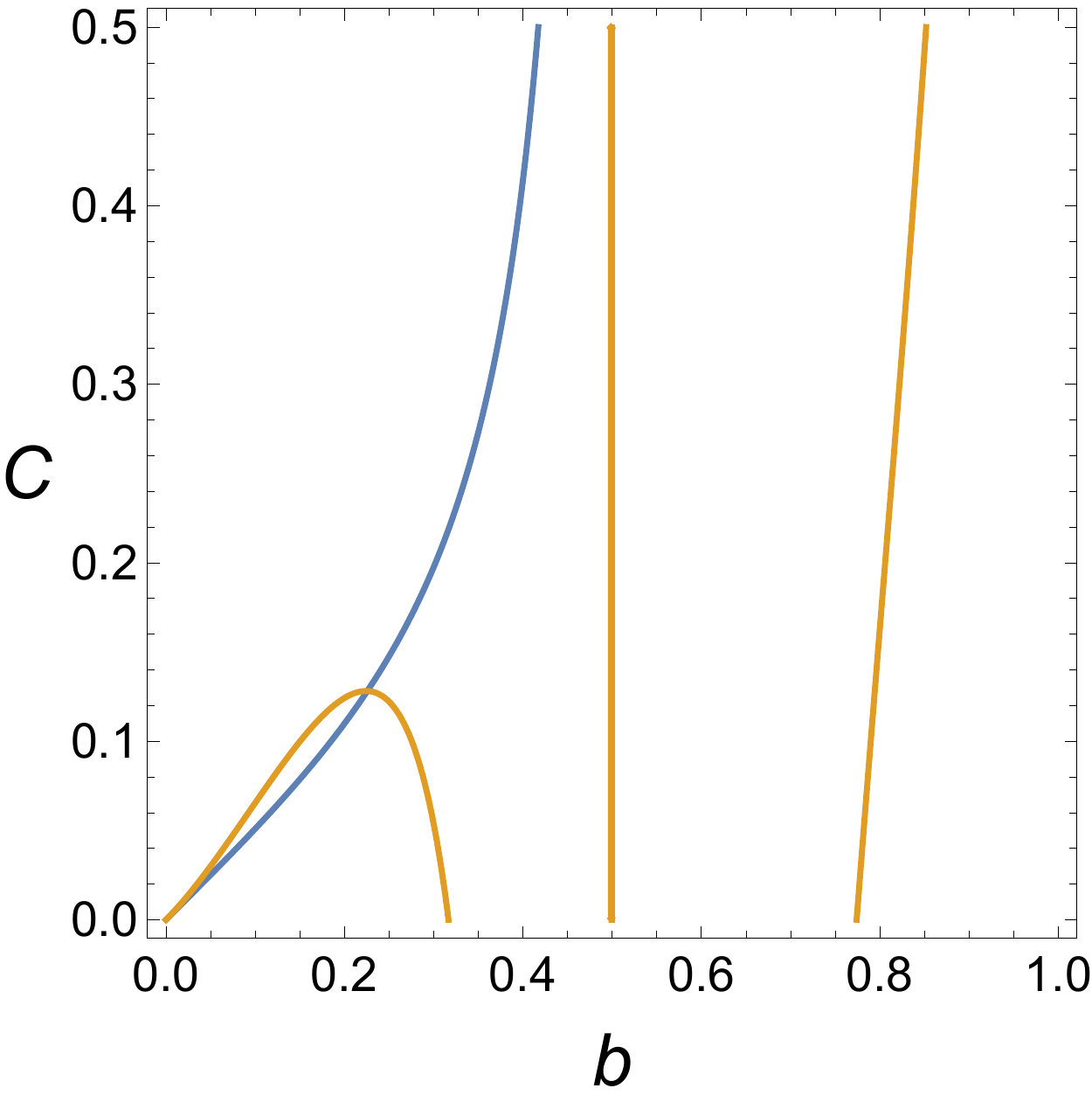}
\caption{Solutions to equations (\ref{lowtemp3}) and (\ref{lowtemp4}) for $b \in [0,1]$.}\label{solns}
\end{figure}

Let us assume that the solution to the saddle point equations (\ref{lowtemp1}) and (\ref{lowtemp2}) takes the form:{
\begin{equation}
Q(\omega)=\frac{\alpha_Q}{|\omega|^a}~,\quad\quad\quad\quad\quad S(\omega)=i\,\alpha_S\frac{\text{sign}(\omega)}{|\omega|^b}~.\label{scalingansatz}
\end{equation}}
These correspond to conformal weights $\Delta_Q = (1-a)/2$ for the scalars and $\Delta_S = (1-b)/2$ for the fermions.\footnote{At finite temperature, we can obtain the expression for conformally invariant correlators by mapping the the line to a circle \cite{Sachdev:2015efa}:
\begin{equation}
{Q}(k) = \alpha_Q \, \frac{\Gamma[\Delta_Q+k]}{\Gamma[1-\Delta_Q+k]}~, \quad\quad
{S}(k) =  i \alpha_S \, \frac{\Gamma[\Delta_S+k+1/2]}{\Gamma[1-\Delta_S+k+1/2]}~, \quad\quad k \in \mathbb{Z}~. 
\end{equation}
For low temperatures and large frequencies these are well approximated by (\ref{scalingansatz}).
}
Due to (\ref{scaling}), we have $a=1-2b$. Plugging (\ref{scalingansatz}) into (\ref{lowtemp1}) and (\ref{lowtemp2}) leads to divergences if we are not careful. One may consider regulating them by analytic continuation in the powers $a$ and $b$. That is, over certain ranges, these integrals will be representations of the Euler $\beta$-function
\begin{equation}
\beta(x,y)\equiv\int_0^1  \frac{dt}{t^{1-x}(1-t)^{1-y}}=\int_0^\infty  \frac{dt}{t^{1-x}(1+t)^{x+y}}=\frac{\Gamma(x)\Gamma(y)}{\Gamma(x+y)}~.
\end{equation}
Without loss of generality, we will take {$\omega>0$} in (\ref{lowtemp1}) and (\ref{lowtemp2}). Let us treat (\ref{lowtemp2}) first for simplicity. First define $\omega'\equiv \tilde{m}\omega $ and $C\equiv\alpha_Q\alpha_S^2\lambda^2/\pi$, then we can write (\ref{lowtemp2}) as
\begin{equation} 
-\frac{1}{2C}=\left(\int_{-\infty}^{-1}d\tilde{m}~\frac{\text{sign}(\tilde{m})}{|1+\tilde{m}|^a|\tilde{m}|^b}+\int_{-1}^{0}d\tilde{m}~\frac{\text{sign}(\tilde{m})}{|1+\tilde{m}|^a|\tilde{m}|^b}+\int_{0}^{\infty}d\tilde{m}~\frac{\text{sign}(\tilde{m})}{|1+\tilde{m}|^a|\tilde{m}|^b}\right)~.
\end{equation}
The integrals may be expressed as Euler-$\beta$ functions and combining everything gives:
\begin{equation}
1=\frac{2C\pi^2\csc^2(\pi b)}{\Gamma(1-2b)\Gamma(b)\Gamma(1+b)}~.
\end{equation}
Let us now treat (\ref{lowtemp1}) and label the first and second terms $A$ and $B$ respectively:{
\begin{align}
1=& ~2\int {d\omega'}\frac{Q(\omega)Q(-\omega-\omega')}{\int {d\omega''}~Q(\omega'')Q(-\omega''-\omega')}-2{\lambda^2} \int \frac{d\omega'}{2\pi} ~Q(\omega)S(\omega') S\left(-\omega-\omega'\right)\nonumber\\
\equiv&~A+B~.
\end{align}}
The term labeled $B$ on the right hand side can be treated in the same way as before. Define {$\omega'\equiv \tilde{m}\omega$} and we obtain
\begin{align} 
B=&-2C\bigg(\int_{-\infty}^{-1}d\tilde{m}~\frac{\text{sign}(\tilde{m})\text{sign}(1+\tilde{m})}{|1+\tilde{m}|^b|\tilde{m}|^b}\nonumber\\
&~~~~~~~~~~~~~~~~~~~+\int_{-1}^{0}d\tilde{m}~\frac{\text{sign}(\tilde{m})\text{sign}(1+\tilde{m})}{|1+\tilde{m}|^b|\tilde{m}|^b}+\int_{0}^{\infty}d\tilde{m}~\frac{\text{sign}(\tilde{m})\text{sign}(1+\tilde{m})}{|1+\tilde{m}|^b|\tilde{m}|^b}\bigg)\nonumber\\
=&~\frac{2\pi\,C \cot\left(\tfrac{\pi b}{2}\right)\sec(\pi b)\Gamma(1-b)}{\Gamma(2-2b)\Gamma(b)}~.
\end{align}
$A$ requires a little more care. Let us first treat the denominator of the integrand{
\begin{equation}
f(\omega')\equiv\int {d\omega''}~Q(\omega'')Q(-\omega''-\omega')~.
\end{equation}}
Treating $f(\omega')$ carefully for positive and negative $\omega'$ we find that it can be regulated to give {
\begin{equation}
f(\omega)=\alpha_Q^2\pi^2|\omega|^{4b-1}\frac{\csc^2(\pi b)\sec(2\pi b)}{2\Gamma(1-2b)^2\Gamma(4b)}~.
\end{equation}}
With this, $A$ is given by {(again defining $\omega'= \tilde{l}\omega$)}
\begin{align}
A=&~\frac{4\Gamma(1-2b)^2\Gamma(4b)}{\pi^2\csc^2(\pi b)\sec(2\pi b)}\bigg(\int_{-\infty}^{-1}d\tilde{l}\frac{1}{|\tilde{l}|^{4b-1}|1+\tilde{l}|^{1-2b}}\nonumber\\
&~~~~~~~~~~~~~~~~~~~~~~~~~~~~~~~~~~~+\int_{-1}^0 d\tilde{l}\frac{1}{|\tilde{l}|^{4b-1}|1+\tilde{l}|^{1-2b}}+\int_0^\infty d\tilde{l}\frac{1}{|\tilde{l}|^{4b-1}|1+\tilde{l}|^{1-2b}}\bigg)\nonumber\\
=&\frac{1-4b}{1-2b}\left(1-\tan^2(\pi b)\right)~.
\end{align}
Putting everything together we find two equations in the unknowns $C$ and $b$:
\begin{align}
1&=\frac{1-4b}{1-2b}\left(1-\tan^2(\pi b)\right)+\frac{2\pi\,C \cot\left(\tfrac{\pi b}{2}\right)\sec(\pi b)\Gamma(1-b)}{\Gamma(2-2b)\Gamma(b)}~,\label{lowtemp3}\\
1&=\frac{2C\pi^2\csc^2(\pi b)}{\Gamma(1-2b)\Gamma(b)\Gamma(1+b)}~.\label{lowtemp4}
\end{align}
We plot the contours that satisfy the equations in Figure \ref{solns}. In order for the Euclidean-time correlators to decay at late times, we require $a<1$ and $b<1$. Since $a=1-2b$, this restricts $0<b<1$.
Notice there is a solution compatible with the equations for $0<b<1$ with $(b,C)\approx(0.226,0.128)$. It is now straightforward to check that our initial assumption (\ref{inequality}) is satisfied for this solution, so long as:{
\begin{equation}
\alpha_S \ll (\omega)^{b-1}~, \quad \quad  \omega^{1/2-2b}/\lambda \ll \alpha_Q \ll \omega^{-(1+2b)}~.
\end{equation}}
The second inequality further implies that {$\omega ^{3/2} \ll  \lambda$}, which at large $\lambda$ allows for this solution to be valid for a parametrically large range of $\omega$.

The solution found above is not isolated since $\alpha_Q$ and $\alpha_S$ remain unfixed but related by {$\alpha_Q = C \pi / ( \alpha_S^2 \lambda^2 )$}. At this point we can compute the on-shell action as a function of $\alpha_S$ and find which value of $\alpha_S$ is a critical point. A somewhat tedious calculation yields:{
\begin{equation}\label{alphanonsusy}
\alpha_S \approx {1.39} \times  \frac{1}{\lambda^{2/3}}~.
\end{equation}}

\subsubsection{Supersymmetric solution}



Recall that for all values of the disorder $\Omega_{\vec{\alpha}}$, the theory is supersymmetric. At zero temperature, supersymmetry relates the correlation function of the fermion and boson in a given supermultiplet as: $S(\tau,\tau') = \partial_{\tau'} Q(\tau,\tau')$. This follows from the supersymmetric transformation rules acting on $\langle \bar{\psi}^{i,a}_{\alpha}(\tau) {\phi}^i_\alpha(\tau') \rangle = 0$. One can check that it holds explicitly to low order in a small $\lambda$ expansion, as discussed in appendix \ref{pert2}. In combination with the scaling symmetry at large $\lambda$, the quantum mechanics becomes super-conformal \cite{Fubini:1984hf}, such that the scaling dimensions of the bosons and fermions differs by one-half. This imposes that  $b=a-1$ which, in combination with (\ref{scaling}), leads to $b=0$ and $a=1$. If, in addition, the supersymmetic ground state preservers scale invariance, the zero temperature momentum space correlators would behave as:{
\begin{equation}\label{scalingansatz2}
{Q(\omega)} = \frac{1}{2\pi} \frac{\alpha_Q}{ |\omega|}~,\quad\quad S(\omega)=  i\,\alpha_S \, {\text{sign}(\omega)}~.
\end{equation}}
Recall that {$\text{sign}(\omega)$} scales like a constant. In addition supersymmetry relates $\alpha_Q = \alpha_S$ which in turn implies that they both decay as $\lambda^{-2/3}$ at large $\lambda$. This is consistent with the behavior of (\ref{alphanonsusy}).

The scaling ansatz (\ref{scalingansatz2}), for which $\Delta_Q = 0$ and $\Delta_S=1/2$, has a Fourier transform back to Euclidean time which is logarithmically divergent. This is reminiscent of logarithmic divergences that appear in conformal field theories which often indicate the presence of a small scaling anomaly \cite{Bzowski:2013sza}. We suspect that this is the case here as well, and the scaling form of (\ref{scalingansatz2})  is approximate. A natural IR cutoff whose role would be to tame this logarithmic divergence is the Fayet-Iliopoulos parameter $\theta$ that appears in the full quiver theory, including vector multiplets, enforcing that the scalars $\phi_\alpha^i$ take values in a compact space. The smallness of $\Delta_Q$ would suggest that the dynamics of the scalars are effectively frozen as compared to those of the fermions whose correlations decay in time. In appendix \ref{fermionM} we discuss a toy model that has fermions with random masses, that shares some of above features.\footnote{With weight $\Delta_S = 1/2$, the low energy theory allows for an $SO(3)$ breaking marginal (at least at large $N$) deformation: $J = \bold{a} \cdot \int d\tau \left( \bar{\psi}_\alpha^{\dot{a}}(\tau) {\boldsymbol{\sigma}}_{\dot{a} a} \psi_\alpha^a(\tau) \right)$. Perhaps this is related to the near horizon of extremally rotating black holes.}


\subsection{Thermodynamics}


Having solved the saddle point equations, we can compute the on-shell action to leading order and obtain the thermodynamic features of the system at low temperature in the paramagnetic phase. For instance the free energy is given by:
\begin{equation}
F[\beta,\lambda] = \frac{1}{\beta} \lim_{n\to0} \frac{S_{\rm eff}[\beta,\lambda]}{n}~.
\end{equation}
It is convenient to scale out the temperature from the functions $Q(u)$ and $S(u)$, where $u=\tau-\tau'$. These can be written as functions of the dimensionless quantities $u/\beta$, $\lambda^2\beta^3$ such that:
\begin{equation}
Q(u;\lambda^2,\beta) = \beta \, \tilde{Q}(u/\beta;\lambda^2 \beta^3)~, \quad S(u;\lambda^2,\beta) = \tilde{S}(u/\beta;\lambda^2 \beta^3)~.
\end{equation}
 Only explicit factors of $\beta$ will play a role when taking derivatives of $\beta F[\beta,\lambda]$, since we are assuming that $Q(u)$ and $S(u)$ take their on-shell values. Thus we can compute, for example, the internal energy $U[\beta,\lambda] = \partial_\beta \left( \beta F[\beta,\lambda] \right)$ as:
\begin{equation}\label{Uint}
\frac{U[\beta,{\lambda}]}{3 N} = - \frac{3}{2\beta} \sum_{k \in \mathbb{Z}} \left[2\pi i \, (k+1/2) \, \tilde{S}(k) + (2\pi k)^2 \tilde{Q}(k) \right]~.
\end{equation}
Notice that when $S(k)$ and $Q(k)$ take their free values, $U[\beta,\lambda]$ vanishes. This had to be the case, since in the absence of any dimensionful parameters, $S_{\rm eff}[\beta]$ must be independent of $\beta$.

For the non-supersymmetric solution $(b,C)\approx(0.226,0.128)$
we find the low temperature result (upon $\zeta$-function regularization of the infinite sums):
\begin{equation}
\frac{U[\beta,{\lambda}]}{3 N}  \approx {2.11} \times \frac{1}{\lambda^{2/3}} \, \left( \frac{1}{\beta} \right)^2~.  
\end{equation}
The internal energy of the system grows quadratically with the temperature giving rise to a specific heat that is itself linear in the temperature. This resembles the universal low temperature behavior for the specific heat of near extremal black holes. A similar situation holds for the supersymmetric solution. At zero temperature, the entropy also has a contribution from the supersymmetric ground state degeneracy which is also extensive in $N$ \cite{Denef:2007vg}.


{Thus, the replica symmetric phase, to leading order in the large $N$ limit, is governed by a gapless low temperature phase. In section \ref{rep} we established that the replica symmetric phase is perturbatively stable. However, it remains an open question whether there is a glassy replica symmetry broken phase in the system. The possibility of such a replica symmetry broken phase and its holographic interpretation (perhaps related to multi-horizon geometries \cite{Anninos:2013mfa}) is extremely interesting. We hope to address this question in the future, employing a numerical analysis.}

\section{Quenched Coulomb branch}\label{Coulomb}

In our treatment up to now, we have ignored the vector multiplet degrees of freedom $\bold{X}_i = \{ \bold{x}_i, \lambda_i , D_i, A_{i}\}$ representing the position degrees of freedom of the wrapped branes in the non-compact $(3+1)$-dimensional Minkowski space-time. In this section we briefly discuss the effect of having quenched and random $\omega_{\vec{\alpha}}$ on the Coulomb branch, upon integrating out the chiral multiplet. The interaction between the two multiplets is dictated by the following action:
\begin{equation}\label{coulombS}
S_{\rm int} = \int d\tau \left[ \left( \bold{x}_{ij}^2 + D_{ij} \right) |\phi_\alpha^{ij}|^2 + \bar{\psi}_\alpha^{ij} \boldsymbol{\sigma} \psi_\alpha^{ij} \cdot \bold{x}_{ij} +   i2\sqrt{2}  \, \text{Im} \, \bar{\phi}^{ij}_\alpha \lambda_{ij} \epsilon \psi^{ij}_\alpha \right]~,
\end{equation}
where $\bold{x}_{ij} = (\bold{x}_i - \bold{x}_j)$ and so on.\footnote{For convenience we have used a slightly different notation in (\ref{coulombS}), where the chiral multiplet is now labeled by two integers, $(ij)$, denoting the particular two branes they connect.} As before, if we consider the $\omega_{\vec{\alpha}}$ to be quenched random variables we can integrate over them and obtain an effective action $S_{\rm eff}$. The new feature is that $S_{\rm eff}$ will also be a functional of the vector multiplet degrees of freedom $\bold{X}_i$. If we are interested in the effective action of $\bold{x}_i$ only, we can set $\lambda_i = 0$. Then, after similar manipulations to those already performed, we obtain a contribution to the effective action:
\begin{multline}\label{seffcoulomb}
\frac{\delta S_{\rm eff}}{n N}=  \int d\tau d\tau'\delta(\tau-\tau') \left[-\delta_{ij}\partial_\tau^2 +\left(\bold{x}_{ij}(\tau)^2 +D_{ij}(\tau) \right) \right]Q^{ij}(\tau,\tau') \\ + \int d\tau d\tau' \delta(\tau-\tau')\left[ \delta_{a\dot{b}}\delta_{ij}\partial_\tau + \boldsymbol{\sigma}_{a\dot{b}} \cdot \bold{x}_{ij}(\tau) \right] S^{ij,a\dot{b}}(\tau,\tau')~.
\end{multline} 
In \cite{Anninos:2013nra} it was shown that in the absence of a superpotential, the effective multi-particle theory of the $\bold{x}_{i}$ contained a low energy $SL(2,\mathbb{R})$ invariant sector upon integrating out the chiral multiplets and taking a large $N$ limit. The scaling dimension of $\bold{x}_i$ in the low energy sector of the Coulomb branch was found to be $\Delta_{\bold{x}} = 1$. In order for the contribution (\ref{seffcoulomb}) to preserve the scale invariance of the (paramagnetic) effective action (\ref{parS1}), the scaling dimension of $S(\tau,\tau')$ would have to vanish. But this is inconsistent with the scaling dimension of $S(\tau,\tau')$ we found in the previous section. In other words, the two $SL(2,\mathbb{R})$ phases of the full quiver theory, the one in the Coulomb branch and the other in the Higgs branch, are distinct. Going from one to the other, which in the gravity limit might be viewed as the fragmented tips in the warped throat merging into a single horizon, resembles an RG flow from one IR fixed point to another. Somewhat interestingly, $\Delta_{\bold{x}}$ is twice the conformal weight of the fermion for the supersymmetric solution \cite{Berkooz:1999iz}.

\section*{Acknowledgements}

It is a pleasure to acknowledge discussions with Chris Beem, Juan Maldacena, David Poland, Dan Roberts, Douglas Stanford and especially Steve Shenker. D.A. acknowledges funding from the NSF. T.A. is supported by the U.S. Department of Energy under grant Contract Number DE-SC0012567 and by NSF grant PHY-0967299. F.D. is supported in part by the U.S. Department of Energy (DOE) under DOE grant DE-SC0011941. All authors were also partly supported by a grant from the John Templeton Foundation. The opinions expressed in this publication are those of the authors and do not necessarily reflect the views of the John Templeton Foundation.


\appendix

\section{Fermions with random masses}\label{fermionM}

We provide a simple purely fermionic model with random masses as an example a solvable model, which we can also solve using the replica trick. The Hamiltonian is given by:
\begin{equation}
H = J_{\alpha\beta} \bar{\psi}_\alpha \psi_\beta~, \quad\quad \alpha = 1,2,\ldots,N~,
\end{equation}
where $J_{\alpha\beta}$ is an $N\times N$ Hermitean matrix drawn from a Gaussian ensemble with variance $J^2/N$. The $2^N$-dimensional Hilbert space can be decomposed into basis vectors built from the state $|0\rangle$ annihilated by all $\psi_\alpha$, where we recall $\{ \bar{\psi}_\alpha, \psi_\beta \} = \delta_{\alpha\beta}$. The basis vectors are
\begin{equation}
 | \alpha_i ; n \rangle = \prod_{i=1}^n \bar{\psi}_{\alpha_i} |0\rangle~, \quad\quad n \in [0,N]~.
 \end{equation}
A useful quantity characterizing the states is the number of particles $n$, which is the eigenvalue of the number operator. For a given $n$ there are $C^N_n$ states. The Hamiltonian becomes block diagonal with $N$ blocks of size $C^N_n \times C^N_n$ with $n=0,1,\ldots,N$.
The corresponding Euclidean action is:
\begin{equation}
S = \int_0^\beta d\tau \left( \bar{\psi}_\alpha \partial_\tau \psi_\alpha - J_{\alpha\beta} \bar{\psi}_\alpha \psi_\beta  \right)~. 
\end{equation}
Since the model is quadratic, it can be solved exactly. For instance, going to thermal frequency space, the exact two-point function is given by:
\begin{equation}
\langle \, \bar{\psi}_\alpha(k) \psi_\beta(-k-1) \, \rangle_J = \left[ i \, \omega_k   \otimes \mathbb{I}_{N\times N}- J \right]_{\alpha\beta}^{-1}~. 
\end{equation}
We can average the two-point function $\langle \, \bar{\psi}_\alpha(k) \psi_\alpha(-k-1) \, \rangle$, with $\alpha$ summed, over the disorder by computing:
\begin{equation}
\int dJ_{\alpha\beta} \, e^{-J_{\alpha\beta} J_{\beta\alpha} N/2J^2} \, \text{tr}  \, \left[ i \, \omega_k \otimes \mathbb{I}_{N\times N}- J \right]_{\alpha\beta}^{-1}~.
\end{equation}
This is a standard exercise in matrix integrals. It is known that the eigenvalue distribution of $J$ with Gaussian weight is the Wigner semicircle distribution \cite{brezin}, hence we must compute (in the $\beta \to \infty$ limit):
\begin{multline}\label{resolvent}
\frac{1}{N} \, \sum_\alpha {\overline{\langle \, \bar{\psi}_\alpha(\omega) \psi_\alpha(-\omega) \, \rangle_J}} \\ = \frac{1}{2\pi J^2} \, \int_{-2J}^{2J} d\lambda \frac{ \sqrt{(2J-\lambda)(2J+\lambda)}}{i\, \omega -\lambda} = \frac{i}{2 J^2} \left(\omega-\text{sign}(\omega)\sqrt{4 J^2+\omega^2}\right)~.
\end{multline}
In the matrix model literature this often referred to as the resolvent. Notice that in the large $J$ limit, the two-point function is approximately given by the sign function just like the large $\tilde{\lambda}$ correlator (\ref{scalingansatz2}) found in the main body of the text. We can also compute the quench averaged free energy :
\begin{equation}
\frac{\overline{F[\beta,J]}}{N} =  -\frac{1}{2\pi \beta J^2}\, \int_{-2J}^{2J} d\lambda \sqrt{(2J-\lambda)(\lambda+2J)} \log \left( 2 \cosh \frac{\beta\lambda}{2} \right)~.
\end{equation}
From this we can derive an expression for the specific heat:
\begin{equation}
\frac{\overline{C[\beta,J]}}{N} = \frac{4}{2\pi J^2 \beta^2} \int_{-J\beta}^{J\beta} du   \sqrt{(J \beta -u)(u+J\beta)} \, u^2 \, \text{sech}^2 u~.
\end{equation}
In the  low temperature, $\beta \to \infty$ limit the specific heat is linear and goes as $\overline{C[\beta,J]} \approx  \pi N / ( 3 J\beta )$.

We can also solve this model using the techniques outlined in the main body of the text. Hence we should compute the effective theory of $Q(\tau,\tau')$. For the paramagnetic ansatz we find:
\begin{equation}
\frac{S_{\rm eff}}{N n} = \text{tr} \, \log Q(\tau,\tau')+\int d\tau d\tau'   \left( \delta(\tau-\tau') \partial_\tau Q(\tau,\tau') + \frac{J^2}{2} Q(\tau,\tau')Q(\tau',\tau) \right)~.
\end{equation}
The momentum space equations are:
\begin{equation}
\frac{1}{Q(\omega)} = i \, \omega  - J^2 Q(\omega)~,
\end{equation}
with solution:
\begin{equation}
Q_\pm(\omega) = \frac{i }{2 J^2} \left(\omega \pm \sqrt{4 J^2  + \omega ^2}  \right)~.
\end{equation}
From the above solutions, we pick the one for which the physical condition $\overline{Q(\omega)} = Q(-\omega)$ holds, which is the same as (\ref{resolvent}). In this language we can also compute the thermodynamic quantities of the model. The internal energy is given by:
{\begin{equation}
\frac{U[\beta,J]}{N} = \frac{1}{\beta} \sum_{k \in \mathbb{Z}} \left[ 2\pi i (k+1/2) \tilde{Q}(k) - 1 \right] \approx \frac{\pi}{6 \beta ^2 J}~,
\end{equation}
where we have taken a low temperature limit and used $\zeta$-function regularization to evaluate the sum in the second equality. The quantity $\tilde{Q}$ is the dimensionless two point function, similar to those that appeared in (\ref{Uint}) and is defined as:
\begin{equation}
\tilde{Q}(k)=\frac{i}{2\,\beta^2\,J^2}\left\lbrace2\pi\left(k+\frac{1}{2}\right)-\text{sign}\left(k+\frac{1}{2}\right)\sqrt{\left(2\pi\left(k+\frac{1}{2}\right)\right)^2+4\beta^2 J^2}\right\rbrace~.
\end{equation}
} As in the quiver model, we find an internal energy proportional to the temperature squared, giving rise to a linear in temperature specific heat $C \approx  \pi N / ( 3 J\beta )$.


%

\section{Perturbative expansion}\label{pert2}

In this appendix we analyze the perturbative expansion in $\lambda$ of equations (\ref{bose2pt}) and (\ref{fermi2pt}). We discuss the solution with all $Q^i(k) \equiv Q(k)$ equal and all $S^{i,a}(k) \equiv S(k)$ equal. To leading order in the small $\lambda$ expansion, the solutions are:
\begin{eqnarray}
Q^i_0(k) &=& \frac{1}{(2\pi k)^2}~, \quad \forall \quad k \neq 0~, \\
Q^i_0(0) &=& \frac{1}{\lambda}~, \\
S^{i,a}_0(k) &=&  \frac{i}{2\pi(k+1/2)} \quad  \forall \quad  k \in \mathbb{Z}~. 
\end{eqnarray}
The next order is found by expanding $Q^i = Q^i_0 +\delta Q^i$ and $S^{i,a} = S_0^{i,a} +\delta S^{i,a}$. Solving for $\delta S^{i,a}$ and $\delta Q^i$, we find:
\begin{equation}
\delta S^{i,a}(k) = -\frac{2\lambda }{(2 \pi)^3}\frac{i}{\left( k+1/2 \right)^3} + \mathcal{O}(\lambda^2)~,
\end{equation}
and
\begin{eqnarray}
\delta Q^i(k) = -\frac{2\lambda}{(2\pi k)^4} + \mathcal{O}(\lambda^2)~ \quad  \forall \quad k \neq 0~.
\end{eqnarray}
Notice that the permutation symmetry between the different node indices is unbroken. Also notice that at large $k$, $\delta S^{i,a}(k) = \left(2\pi i  k\right)  \delta Q^{i}(k)$ which is consistent with unbroken supersymmetry.



\end{document}